\documentclass[reprint,
amsmath,amssymb,
 aps,
prl,
]{revtex4-2}

\usepackage{graphicx}\usepackage{dcolumn}\usepackage{bm}

\usepackage{amsmath}
\usepackage{amssymb}
\usepackage{booktabs}
\usepackage{hyperref}
\hypersetup{
    colorlinks,
    citecolor=blue,
    linkcolor=blue,
    urlcolor=blue
}

\usepackage[whole]{bxcjkjatype}
\usepackage[normalem]{ulem}
\usepackage{color}
\usepackage{soul}
\definecolor{americanrose}{rgb}{1.0, 0.01, 0.24}
\definecolor{electricpurple}{rgb}{0.75, 0.0, 1.0}
\definecolor{vividgreen}{rgb}{0.0, 0.6, 0.0}

\newcommand{\Eref}[1]{Eq.~\eqref{#1}}

\newcommand{\Fref}[1]{Fig.~\ref{#1}}
\newcommand{\FFref}[1]{Figure~\ref{#1}}

\begin{document}

\preprint{APS/123-QED}

\title{Kinetic Uncertainty Relation in Collective Dissipative Quantum Many-Body Systems}

\author{Hayato Yunoki}
 \email{yunoki@biom.t.u-tokyo.ac.jp}

\affiliation{
 Department of Information and Communication Engineering,\\
 Graduate School of Information Science and Technology,\\
 The University of Tokyo, Tokyo 113-8656, Japan
}

\author{Yoshihiko Hasegawa}\email{hasegawa@biom.t.u-tokyo.ac.jp}

\affiliation{Department of Electrical Engineering and Information Systems, Graduate School of Engineering, The University of Tokyo,
Tokyo 113-8656, Japan}

\begin{abstract}
Attaining the ultimate precision remains a central objective in the engineering of nanoscale systems and the investigation of nonequilibrium processes. While thermodynamic and kinetic uncertainty relations establish fundamental precision bounds, prior derivations in the quantum regime have remained confined to single-body systems. Consequently, the ultimate precision limits for interacting many-body systems have been unknown. In this Letter, we analytically formulate a kinetic uncertainty relation for a many-body system undergoing collective dissipation, a paradigmatic model of boundary time crystals. By applying a mean-field approximation, we derive lower bounds for relative fluctuations expressed in terms of clear physical quantities. Our analysis identifies a cooperative enhancement mechanism, demonstrating that collective interactions allow the precision to scale with the number of particles. We validate these findings through numerical simulations across the stationary, critical, and boundary time crystal phases. Our work presents the first theoretical description of precision bounds in collective dissipative quantum many-body systems for an arbitrary particle number $N$, providing a solid foundation for designing future quantum technologies that exploit many-body phenomena.
\end{abstract} 
\maketitle

\paragraph{Introduction.---}
Achieving extreme precision in physical processes is a primary goal in designing nanoscale devices and exploring nonequilibrium dynamics. Recent developments in stochastic thermodynamics have revealed a universal trade-off governing such processes, demonstrating that achieving higher precision inevitably demands a correspondingly greater thermodynamic cost. This principle determines the maximum achievable precision for a given system and quantifies how far its current operation lies from the ultimate theoretical boundary. The inequalities providing these fundamental limits on stochastic processes are known as thermodynamic uncertainty relations (TURs) \cite{baratoThermodynamicUncertaintyRelation2015, gingrichDissipationBoundsAll2016, horowitzThermodynamicUncertaintyRelations2020}. 
For classical Markov jump processes, the relative fluctuation of a time-integrated current $Q$ satisfies
\begin{equation}\label{classical_tur}
    \frac{\text{Var}[Q]}{\langle Q\rangle}\geq\frac{1}{\mathcal{C}},
\end{equation}
where $\mathrm{Var}[Q]$ and $\langle Q \rangle$ denote the variance and the expectation value of the current, respectively. Here, the cost factor is given by $\mathcal{C}=\Sigma/2$, with $\Sigma$ representing the total entropy production. Furthermore, for an arbitrary time-integrated observable, a complementary inequality known as the kinetic uncertainty relation (KUR) holds when the cost is replaced by the dynamical activity $\mathcal{C}=\mathcal{A}_c$, which denotes the total number of jumps \cite{garrahanSimpleBoundsFluctuations2017, diterlizziKineticUncertaintyRelation2019}. 

Recent works have extended these frameworks to the quantum regime, discovering that quantum systems can violate the classical bound in \Eref{classical_tur} \cite{erkerAutonomousQuantumClocks2017, guarnieriThermodynamicsPrecisionQuantum2019, carolloUnravelingLargeDeviation2019, hasegawaQuantumThermodynamicUncertainty2020, hasegawaThermodynamicUncertaintyRelation2021, hasegawaIrreversibilityLoschmidtEcho2021, millerThermodynamicUncertaintyRelation2021, hasegawaThermodynamicUncertaintyRelation2022, vanvuThermodynamicsPrecisionMarkovian2022, prechRoleQuantumCoherence2025, yunokiQuantumSpeedLimit2025a, ishidaQuantumComputerBasedVerification2025, honmaInformationthermodynamicBoundsPrecision2026}. Specifically, purely quantum effects such as coherence can suppress dynamical fluctuations, yielding higher precision than classically permitted. Formulating these quantum precision limits directly dictates the ultimate stability of quantum devices such as quantum clocks and quantum batteries. 

As the pursuit of these advanced quantum technologies moves toward macroscopic scales, interacting quantum many-body systems naturally form their physical backbone. These systems exhibit exotic fundamental phenomena, such as macroscopic phase transitions driven by collective effects, and simultaneously underpin the development of complex solid-state architectures and future quantum information technologies \cite{andersonMoreDifferentBroken1972}. 

While classical TURs were successfully derived for many-body models \cite{shpielbergThermodynamicUncertaintyRelations2021, koyukThermodynamicUncertaintyRelation2022}, prior investigations into quantum precision bounds remained strictly confined to single-body systems. Formulating these precision limits for the quantum many-body regime therefore emerges as a crucial theoretical objective. Conventional theoretical frameworks for TUR and KUR in quantum systems have primarily analyzed the precision of individual quantum trajectories within standard Markovian dissipative setups. How genuine many-body effects, specifically collective dissipation, redefine these fundamental precision boundaries remains completely unknown. For instance, a recent study evaluated the precision limit of the first-passage time under continuous homodyne measurement in a system with collective dissipation \cite{singhQuantumThermodynamicsLimit2026}. However, their exact calculations were limited to very small systems with $N=2$ and $3$. Consequently, TUR and KUR for quantum many-body systems with an arbitrary particle number $N$ remain completely unexplored. Recent experimental advancements successfully extracted information directly from collectively dissipating quantum many-body systems \cite{ferioliNonequilibriumSuperradiantPhase2023}. This emerging observational capability motivates the theoretical formulation of precision bounds for such setups.

A paradigmatic phenomenon governed by such collective dissipation and coherent driving is the boundary time crystal (BTC) \cite{ieminiBoundaryTimeCrystals2018}. Wilczek originally proposed the concept of a time crystal as a physical state that spontaneously breaks time-translation symmetry \cite{wilczekQuantumTimeCrystals2012}. Early theoretical investigations primarily focused on the discrete breaking of time-translation symmetry in periodically driven setups \cite{sachaModelingSpontaneousBreaking2015, elseFloquetTimeCrystals2016, khemaniPhaseStructureDriven2016, russomannoFloquetTimeCrystal2017, gongDiscreteTimeCrystallineOrder2018, suraceFloquetTimeCrystals2019, tuqueroDissipativeTimeCrystal2022}, and numerous experiments subsequently confirmed the existence of these discrete time crystals \cite{choiObservationDiscreteTimecrystalline2017, zhangObservationDiscreteTime2017, rovnyObservationDiscreteTimeCrystalSignatures2018, smitsObservationSpaceTimeCrystal2018, kyprianidisObservationPrethermalDiscrete2021, kesslerObservationDissipativeTime2021, freyRealizationDiscreteTime2022, miTimecrystallineEigenstateOrder2022}. 
Subsequently, researchers demonstrated that continuous time-translation symmetry can spontaneously break in dissipative quantum many-body systems, and such a nonequilibrium phase is termed the BTC. In this phase, order parameters exhibit persistent self-sustained oscillations in the thermodynamic limit.

Following its initial formulation, the BTC concept was generalized, and its emergence across various models and physical platforms was confirmed both theoretically and experimentally \cite{tuckerShatteredTimeCan2018, bucaNonstationaryCoherentQuantum2019, lledoDissipativeTimeCrystal2020, seiboldDissipativeTimeCrystal2020, bookerNonstationarityDissipativeTime2020, prazeresBoundaryTimeCrystals2021, piccittoSymmetriesConservedQuantities2021, carolloExactSolutionBoundary2022, hajdusekSeedingCrystallizationTime2022, doi:10.1126/science.abo3382, krishnaMeasurementInducedContinuousTime2023, wangBoundaryTimeCrystals2025, russoQuantumDissipativeContinuous2025}. Consequently, the BTC has been extensively studied in recent years as a fascinating phenomenon in open quantum many-body systems. The paradigmatic model of a BTC consists of an ensemble of spin-1/2 particles subject to superradiant decay induced by a Markovian bath, and its dynamics are described by \Eref{lindblad}. The competition between the coherent driving and the collective dissipation gives rise to distinct nonequilibrium regimes, including stationary and oscillatory phases.

Recent works have actively explored the application of time crystals induced by collective dissipation to quantum metrology \cite{montenegroQuantumMetrologyBoundary2023, cabotContinuousSensingParameter2024} or quantum clock \cite{viottiQuantumTimeCrystal2026}, which is one of the most prominent applications of the TUR or KUR. Because the BTC is an inherently open system, it allows for continuous measurement \cite{landiCurrentFluctuationsOpen2024}, prompting detailed studies on its quantum trajectories. Ref.~\cite{cabotQuantumTrajectoriesDissipative2023} pointed out that the dynamical phases of the BTC can be precisely characterized using accessible experimental records such as photon count signals and homodyne currents.

\begin{figure}[t]
    \centering
    \includegraphics[width=\linewidth]{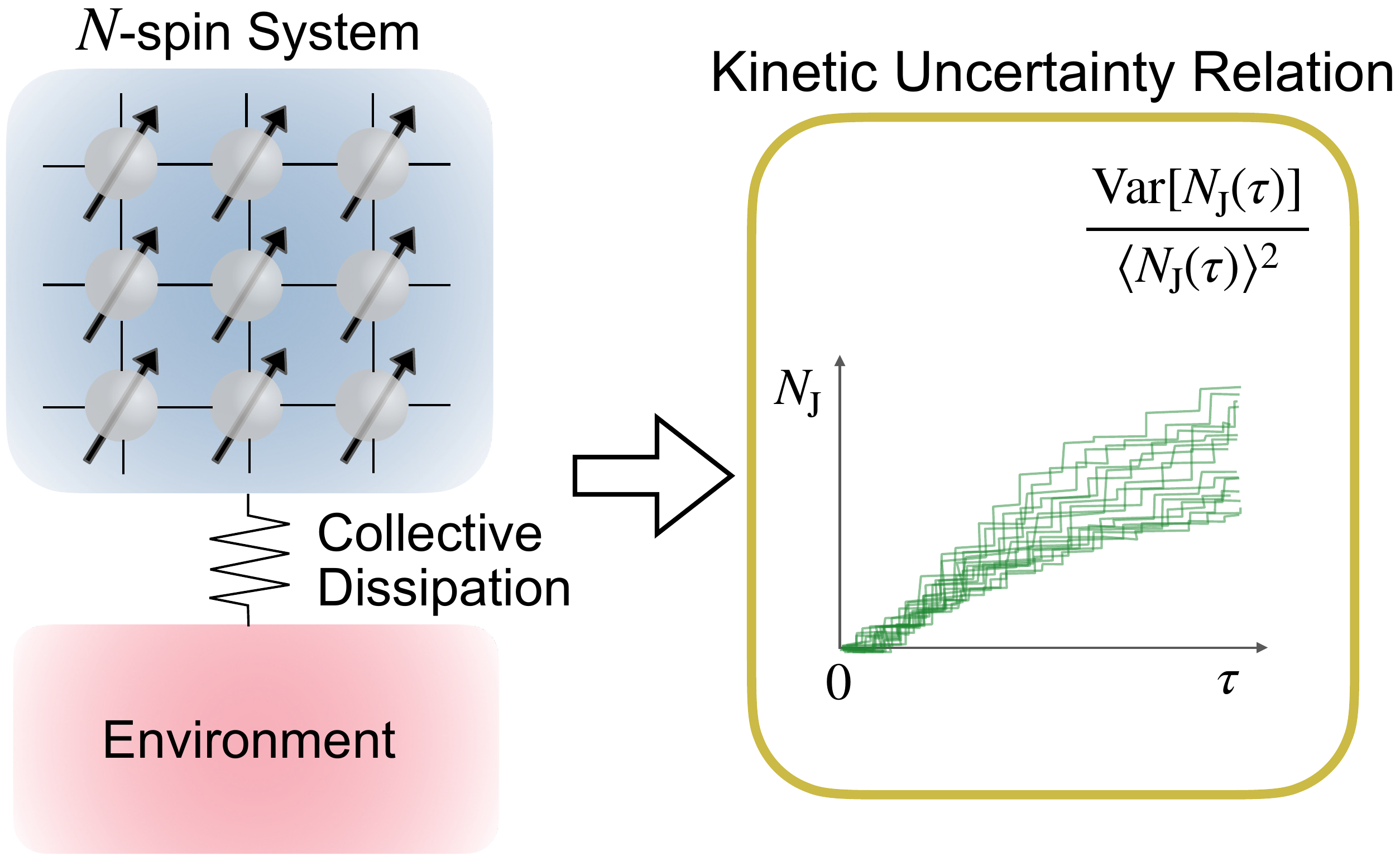}
    \caption{Schematic illustration of the kinetic uncertainty relation in a collective dissipative quantum many-body system. The setup consists of an ensemble of $N$ interacting spins subject to collective dissipation into a Markovian environment, which serves as a paradigmatic model for boundary time crystals. Continuous measurement yields individual stochastic quantum jump trajectories. The quantity $N_\mathrm{J}(\tau)$ represents the accumulated number of jumps over an observation time $\tau$. We derive fundamental precision limit of the relative fluctuations, which is governed by the kinetic uncertainty relation.}
    \label{fig:schematic_intro}
\end{figure}

In this Letter, we derive a KUR for quantum trajectories governed by the collective dissipative dynamics in \Eref{lindblad}, which serves as a paradigmatic model of the BTC (\Fref{fig:schematic_intro}). This formulation allows us to clarify how genuine many-body effects dictate the fundamental limits of precision. Although most existing TUR and KUR in quantum systems establish analogous limits, the resulting inequalities typically depend on the density matrix or abstract quantities lacking clear physical interpretation. We overcome this fundamental difficulty by utilizing the many-body mean-field approximation \cite{benattiQuantumSpinChain2018, carolloExactnessMeanFieldEquations2021}, successfully bounding the precision with physically transparent observables. Furthermore, we validate our analytically derived inequality through numerical simulations. We investigate how the actual precision and its theoretical lower bound behave across the stationary, critical, and BTC phases. Finally, we analyze the scaling of the precision with respect to the particle number $N$ to uncover the system-size dependence unique to many-body systems.

 \paragraph{Methods.---}
As a paradigmatic system featuring collective dissipation, we consider an ensemble of $N$ spin-$1/2$ particles subject to coherent resonant driving. The dynamics of this open quantum system are governed by a Markovian master equation in the Lindblad form \cite{goriniCompletelyPositiveDynamical1976a, lindbladGeneratorsQuantumDynamical1976}. This model has been widely studied in the context of cooperative resonance fluorescence \cite{agarwalCollectiveAtomicEffects1977, narducciTransientSteadystateBehavior1978, drummondVolterraCyclesCooperative1978, carmichaelAnalyticalNumericalResults1980} and cooperative emission in cavities \cite{10.1143/PTPS.64.307, puriExactSteadystateDensity1979, wallsCooperativeFluorescenceCoherently1980, schneiderEntanglementSteadyState2002, hannukainenDissipationDrivenQuantum2018}, and it accurately describes a variety of experimental setups \cite{shankarSteadystateSpinSynchronization2017b, doi:10.1126/science.aar3102}. Recently, this setup has also been extensively investigated as the standard model for studying the BTC \cite{ieminiBoundaryTimeCrystals2018}.

To describe the system of $N$ particles, we introduce the collective spin operators $S_\alpha = \frac{1}{2} \sum_{i=1}^N \sigma_\alpha^{(i)}$ ($\alpha = x,y,z$) and the corresponding ladder operators $S_\pm = S_x \pm i S_y$. The time evolution of the system's density matrix $\rho$ is governed by the master equation $\dot{\rho} = \mathcal{L}\rho$. The Liouvillian superoperator $\mathcal{L}$ is defined as
\begin{equation}\label{lindblad}
    \mathcal{L}\rho \equiv -i[\omega S_x, \rho] + \frac{2\kappa}{N} \left( S_- \rho S_+ - \frac{1}{2} \{S_+ S_-, \rho\} \right).
\end{equation}
The first term on the right-hand side describes the unitary evolution generated by the Hamiltonian $H=\omega S_x$, where $\omega$ is the Rabi frequency. The second term represents the collective dissipation characterized by the jump operator $L=S_-$. The parameter $2\kappa/N$ dictates the decay rate. The explicit $1/N$ scaling of the rate is necessary to ensure that the dynamics remain well-defined in the thermodynamic limit \cite{ieminiBoundaryTimeCrystals2018, benattiQuantumSpinChain2018}.

To analyze the macroscopic behavior of the system, we define the magnetization components $m_\alpha \equiv \langle S_\alpha\rangle/(N/2)$. By invoking the clustering assumption, which posits that correlations are negligible, the macroscopic operators can be represented as sums over the identity \cite{lanfordObservablesInfinityStates1969, bratteliOperatorAlgebrasQuantum1997, benattiQuantumSpinChain2018, strocchiSymmetryBreaking2021}. In other words, the operators converge to their expectation values, $\lim_{N\rightarrow\infty} S_\alpha/(N/2)=m_\alpha$. Consequently, in the thermodynamic limit ($N \to \infty$), the dynamics of the magnetization are captured by the following set of mean-field equations \cite{ieminiBoundaryTimeCrystals2018, carolloExactSolutionBoundary2022}:
\begin{equation}\label{mean_field_equation}
\begin{split}
    \dot{m}_x &= \kappa m_x m_z,\\
    \dot{m}_y &= -\omega m_z + \kappa m_y m_z, \\
    \dot{m}_z &= \omega m_y - \kappa (1 - m_z^2).
\end{split}
\end{equation}
These mean-field equations conserve the total angular momentum, satisfying $m_x^2 + m_y^2 + m_z^2 = 1$.

In the thermodynamic limit, the interplay between coherent driving and collective dissipation induces a nonequilibrium phase transition at the critical point $\omega/\kappa=1$ \cite{ieminiBoundaryTimeCrystals2018}. Depending on the parameter regime, the system exhibits two distinct dynamical phases, which are a stationary phase ($\omega/\kappa<1$) and an oscillatory phase ($\omega/\kappa>1$). In the oscillatory phase, a macroscopic order parameter exhibits sustained periodic oscillations. While these oscillations eventually damp out for a finite particle number $N$, the macroscopic order parameter maintains persistent oscillation as the system reaches the thermodynamic limit. This sustained dynamic behavior represents the spontaneous breaking of continuous time-translation symmetry. The resulting oscillatory phase is defined as the BTC phase.

While the Lindblad master equation dictates the average time evolution of the density matrix, the deterministic dynamics can be decomposed into individual stochastic trajectories. This decomposition is known as unraveling, and it establishes a direct correspondence with a continuous measurement process  \cite{landiCurrentFluctuationsOpen2024}. Because the present model describes an open quantum system, we can consider the continuous measurement of quantum trajectories \cite{cabotQuantumTrajectoriesDissipative2023, cabotContinuousSensingParameter2024}. Depending on the specific choice of the continuous measurement scheme, an infinite number of distinct unraveling strategies exist for a given Lindblad equation. The most paradigmatic approach is the quantum jump trajectory, which physically corresponds to direct photon counting or jump measurements.

To clarify the connection between the master equation and individual trajectories, we express the infinitesimal time evolution of the density matrix in the Kraus representation
\begin{equation}
    \rho(t+dt)=M_0\rho(t)M_0^\dagger+M_1\rho(t)M_1^\dagger.
\end{equation}
This formulation demonstrates that the deterministic Lindblad dynamics arise from the summation of two distinct physical processes described by the Kraus operators
\begin{equation}
    M_0=I-iH_\text{eff}dt, M_1=\sqrt{\frac{2\kappa}{N}dt}S_-,
\end{equation}
where $I$ is the identity operator and $H_\text{eff}=H-i\frac{\kappa}{N}S_+ S_-$ represents the non-Hermitian effective Hamiltonian. The first term in the Kraus representation corresponds to the evolution when no jump is detected, while the second term accounts for the occurrence of a discrete quantum jump. A jump associated with the operator $M_1$ occurs in the time interval $dt$ with a probability $p_1(t)=\text{Tr}[M_1\rho_c(t)M_1^\dagger]$, where $\rho_c(t)$ is the conditional density matrix of an individual trajectory. Notably, the no-jump evolution governed by $M_0$	is non-unitary because the absence of a detected jump still provides information and continuously updates the knowledge of the state.

We can introduce a random variable $dN_\mathrm{J}(t)$ that takes the value of 1 when a jump is detected and 0 otherwise. Because the time step $dt$ is infinitesimal, the probability of observing more than one jump is negligible, which leads to $dN_\mathrm{J}(t)^2=dN_\mathrm{J}(t)$ and $dN_\mathrm{J}(t)dt=0$. The stochastic evolution of the conditional state is then described by the stochastic master equation
\begin{equation}
\begin{split}
        d\rho_c=&dt\mathcal{L}\rho_c\\
        &+\left(\frac{S_-\rho_cS_+}{\text{Tr}[S_-\rho_cS_+]} - \rho_c\right)\left(dN_\mathrm{J}(t) - dt\frac{2\kappa}{N}\text{Tr}[S_-\rho_cS_+]\right).
\end{split}
\end{equation}
Taking the ensemble average over all possible stochastic sequences recovers the deterministic Lindblad dynamics, $\mathbb{E}[d\rho_c] = \dot{\rho}dt$.

By tracking these individual trajectories, we can define stochastic output currents based on the occurrence of the jumps. A fluctuating current is constructed as the rate of change of a counting variable $dN_\mathrm{J}(t)/dt$. The time-integrated observable corresponding to the total number of detected jumps over an observation time $\tau$ is defined as
\begin{equation}\label{observable}
    N_\mathrm{J}(\tau)=\int_0^\tau dN_\mathrm{J}(t).
\end{equation}
This accumulated jump count fluctuates from one trajectory to another and serves as the fundamental stochastic quantity for our analysis. Both the expectation value and the variance of this integrated current represent the standard observables of interest when investigating thermodynamic and kinetic uncertainty relations in the quantum regime. \paragraph{Results.---}
We derive a fundamental inequality that establishes the precision limit of the current fluctuations for the model introduced in \Eref{lindblad}. This model represents a fundamental framework for many-body open quantum systems featuring collective dissipation. Furthermore, in the limit of a large particle number, the system exhibits the BTC, which represents a crucial dynamical phenomenon unique to many-body systems. To capture the many-body properties, we consider the regime where the particle number $N$ is sufficiently large and employ a mean-field approximation \cite{benattiQuantumSpinChain2018, carolloExactnessMeanFieldEquations2021}. Notably, when our many-body formulation is reduced to a single-body scenario, the inequality reproduces the conventional KUR known in prior studies.

We first map the continuous measurement process, including both the quantum jump trajectories of the system and the corresponding quantum jump event records, onto a continuous matrix product state (cMPS) $|\Psi(\tau)\rangle$ \cite{verstraeteContinuousMatrixProduct2010, osborneHolographicQuantumStates2010}. This formalism serves as a standard approach for deriving thermodynamic and kinetic uncertainty relations in open quantum systems \cite{hasegawaQuantumThermodynamicUncertainty2020, hasegawaUnifyingSpeedLimit2023, hasegawaIrreversibilityLoschmidtEcho2021, hasegawaThermodynamicUncertaintyRelation2022, vanvuFundamentalBoundsPrecision2025, prechRoleQuantumCoherence2025, honmaInformationthermodynamicBoundsPrecision2026}. To intuitively understand this mapping, we divide the total trajectory time $\tau$ into discrete intervals of length $dt$. By sequentially applying the Kraus operators $M_{r_k}$ along with the corresponding environmental basis states $|r_k\rangle$ at each time step $k$, we construct a matrix product state (MPS)
\begin{equation}
    |\Psi(\tau)\rangle=\sum_{r_1, ..., r_{N_\tau}}(M_{r_{N_\tau}}, ...M_{r_1}|\psi_0\rangle)\otimes |r_1, ..., r_{N_\tau}\rangle,
\end{equation}
where $N_\tau=\tau/dt$ denotes the total number of steps and $|\psi_0\rangle$ is the initial state of the system. In this expression, the sequence of indices $(r_1, ..., r_{N_\tau})$ records the sequence of the measurement outcomes within the environment degrees of freedom. Taking the continuum limit $dt\rightarrow 0$ transforms this MPS representation into the cMPS. The resulting $|\Psi(\tau)\rangle$ describes the entire history of the open system's dynamics.

We introduce a parameter $\theta$ to scale the Hamiltonian and the jump operator as
\begin{equation}
    H(\theta)=(1+\theta)S_x, L(\theta)=\sqrt{1+\theta}S_-.
\end{equation}
This specific choice of scaling ensures that the overall timescale of the dynamics governed by the Lindblad equation in \Eref{lindblad} is uniformly multiplied by a factor of $1+\theta$. By substituting these parameterized operators into the construction procedure, we obtain a cMPS $|\Psi(\tau;\theta)\rangle$ that reflects the temporal perturbation. Because this representation encodes the stochastic trajectories as a pure state, the quantum Fisher information $J(\theta)$ \cite{meyerFisherInformationNoisy2021} is determined by
\begin{equation}
        J(\theta)= 4\left[\langle\partial_\theta\Psi(\tau; \theta)|\partial_\theta\Psi(\tau; \theta)\rangle-|\langle\partial_\theta\Psi(\tau; \theta)|\Psi(\tau; \theta)\rangle|^2\right].
\end{equation}
We then consider a parameter estimation problem for $\theta$ using the measurement record of the time-integrated observable $N_\mathrm{J}(\tau)$ in \Eref{observable}. Within the framework of quantum estimation theory, the following quantum Cram\'er-Rao inequality \cite{hottaQuantumEstimationLocal2004} holds 
\begin{equation}\label{cramer}
    \frac{\mathrm{Var}_\theta[N_\mathrm{J}(\tau)]}{(\partial_\theta\langle N_\mathrm{J}(\tau)\rangle_\theta)^2}\geq\frac{1}{J(\theta)},
\end{equation}
where the subscript $\theta$ signifies that the variance and the expectation value are calculated under the perturbed dynamics.

We now evaluate this inequality at $\theta=0$. The variance naturally reduces to the fluctuation of the unperturbed original dynamics, which is expressed as
\begin{equation}\label{var}
    \text{Var}_{\theta=0}[N_\mathrm{J}(\tau)]=\text{Var}[N_\mathrm{J}(\tau)].
\end{equation}
To evaluate the expectation value term, we reparameterize the time scaling by introducing the relation $1+\theta=t/\tau$. Under this substitution, $\theta=0$ corresponds to $t=\tau$. The derivative with respect to $\theta$ is then transformed into the derivative with respect to $t$, yielding the relation
\begin{equation}\label{partial_expecation}
    \partial_\theta\langle N_\mathrm{J}(\tau)\rangle_\theta|_{\theta=0}=\tau\partial_\tau\langle N_\mathrm{J}(\tau)\rangle.
\end{equation}
Furthermore, in the steady state, the expectation value of the time-integrated observable grows linearly with time, which implies $\tau\partial_\tau\langle N_\mathrm{J}(\tau)\rangle = \langle N_\mathrm{J}(\tau)\rangle$. Thus, \Eref{partial_expecation} simplifies to
\begin{equation}\label{partial_expecation_ss}
    \partial_\theta\langle N_\mathrm{J}(\tau)\rangle_\theta|_{\theta=0} = \langle N_\mathrm{J}(\tau)\rangle.
\end{equation}

We next evaluate the quantum Fisher information at the unperturbed limit $J(\theta=0)$. In the Ref~\cite{hasegawaUnifyingSpeedLimit2023}, this quantity is recognized as the quantum dynamical activity. It frequently emerges as the central physical quantity governing the fundamental bounds on both the precision and the speed of Markovian open quantum systems. Under the assumption of a sufficiently large particle number $N$, we analytically calculate this many-body quantum dynamical activity $B_\text{mb}(\tau)$. The detailed derivation is provided in the Supplemental Material \cite{SM_note}, and the resulting expression is explicitly given by
\begin{equation}
    \begin{aligned}\label{B_mb}
&B_\text{mb}(\tau) = \frac{\kappa N}{2} \int_0^\tau dt \left( 1 - m_z(t)^2 \right)\\
&+ 2 N \omega \int_0^\tau ds_1 \int_0^{s_1} ds_2 \\
&\Biggl[ \omega \sum_{\alpha \in \{x,y,z\}} U_{x\alpha}(s_1, s_2) \left( \delta_{x\alpha} - m_x(s_2)m_\alpha(s_2) \right) \\
&+ \kappa m_z(s_2) \left( m_y(s_2) U_{xx}(s_1, s_2) - m_x(s_2) U_{xy}(s_1, s_2) \right) \Biggr].
\end{aligned}
\end{equation}
Here, $U(s_1, s_2)$ is the evolution matrix defined as
\begin{equation}
    U(s_1, s_2) = \mathcal{T} \exp \left( \int_{s_2}^{s_1} dt K(t) \right),
\end{equation}
where $\mathcal{T}$ denotes the time-ordering operator, and the generator $K(t)$ is given by
\begin{equation} 
K(t) \equiv \begin{pmatrix}
\kappa m_z(t) & 0 & \kappa m_x(t) \\
0 & \kappa m_z(t) & \kappa m_y(t) - \omega \\
-2\kappa m_x(t) & -2\kappa m_y(t) + \omega & 0
\end{pmatrix}.
\end{equation}
According to Refs.~\cite{nakajimaSymmetriclogarithmicderivativeFisherInformation2023, nishiyamaExactSolutionQuantum2024}, which evaluated $J(\theta=0)$ for the Lindblad equation in a single-body system, the first term on the right-hand side of $B_\text{mb}(\tau)$ corresponds to the classical dynamical activity, while the second term originates from the effects of quantum coherence.

To gain a more transparent physical interpretation of this result, we introduce an upper bound for $B_{\text{mb}}(\tau)$, denoted as $B_\text{mb}^\text{up}(\tau)$. The detailed derivation of this bound is also provided in the Supplemental Material \cite{SM_note}. This upper bound is expressed as
\begin{equation}
\begin{split}\label{B_mb_up}
    B_\text{mb}^\text{up}(\tau) = & \frac{\kappa N}{2} \int_0^\tau dt \left( 1 - m_z(t)^2 \right)  \\
    &+  2N \left( \int_0^\tau ds_1 \mathcal{F}_S(s_1) \right) \left( \int_0^\tau ds_2 \mathcal{F}_{\mathrm{eff}}(s_2) \right) ,
\end{split}
\end{equation}
where
\begin{equation}
    \mathcal{F}_S(s) = \omega \sqrt{1 - m_x(s)^2},
\end{equation}
\begin{equation}
    \mathcal{F}_{\mathrm{eff}}(s) = \omega \sqrt{1 - m_x(s)^2} + \kappa |m_z(s)| \sqrt{1 - m_z(s)^2}.
\end{equation}
In this representation, $\mathcal{F}_S(s)$ and $\mathcal{F}_{\mathrm{eff}}(s)$ capture the quantum fluctuations driven by the system Hamiltonian and the effective Hamiltonian, respectively. A direct numerical comparison between the exact quantum Fisher information $J(0)$ and our analytical expressions $B_{\mathrm{mb}}(\tau)$ and $B_{\mathrm{mb}}^{\mathrm{ub}}(\tau)$, derived under the large-$N$ and mean-field approximation, is provided in the Supplemental Material \cite{SM_note}. These numerical results confirm that $B_{\mathrm{mb}}(\tau)$ accurately reproduces $J(0)$ within the validity of the approximation, and verify that $B_{\mathrm{mb}}^{\mathrm{ub}}(\tau)$ serves as an upper bound to $B_{\mathrm{mb}}(\tau)$.

By evaluating the quantum Cram\'er-Rao inequality in \Eref{cramer} at $\theta=0$ using Eqs.~(\ref{var}) and (\ref{partial_expecation}), and incorporating the relation $J(0) = B_\text{mb}(\tau) \leq B_\text{mb}^\text{up}(\tau)$, we establish the following fundamental bounds:
\begin{equation}\label{our_kur}
    \frac{\text{Var}[N_\mathrm{J}(\tau)]}{\tau^2(\partial_\tau\langle N_\mathrm{J}(\tau)\rangle)^2}\geq \frac{1}{B_\text{mb}(\tau)}\geq\frac{1}{B_\text{mb}^\text{up}(\tau)}.
\end{equation}
This is our main result, representing the KUR for the many-body model introduced in \Eref{lindblad}. Particularly in the steady state, applying \Eref{partial_expecation_ss} simplifies the denominator, reducing the inequality to a bound on the relative fluctuation:
\begin{equation}
    \frac{\text{Var}[N_\mathrm{J}(\tau)]}{\langle N_\mathrm{J}(\tau)\rangle^2}\geq \frac{1}{B_\text{mb}(\tau)}\geq\frac{1}{B_\text{mb}^\text{up}(\tau)}.
\end{equation}

The explicit forms of $B_{\mathrm{mb}}(\tau)$ and $B_{\mathrm{mb}}^{\mathrm{up}}(\tau)$ provide an intuitive physical interpretation of how many-body dynamics govern the precision limit. The first term represents the classical dynamical activity, which directly corresponds to the expected magnitude of quantum jumps induced by the collective dissipation $S_-$. When the magnetization lies in the transverse plane ($m_z \approx 0$), the transverse phases of the individual spins are highly synchronized. This phase alignment triggers strong interference akin to Dicke superradiance. This cooperative phenomenon amplifies the jump probability, yielding a massive increase in statistical event counts that reduces the uncertainty. The second term embodies the quantum contribution originating from the coherent dynamics driven by the Hamiltonian along the $x$-axis $S_x$. As seen in $\mathcal{F}_S(s)$ and $\mathcal{F}_{\mathrm{eff}}(s)$, this contribution is quantified by the factor $\sqrt{1 - m_x(s)^2}$, which measures the magnitude of the spin components in the orthogonal $y$-$z$ plane. Because the rotational drive $\omega$ acts exclusively on these transverse components, this factor precisely captures the speed of the quantum state change in the orthogonal direction. These coherent dynamics thus function as an additional source of quantum dynamical activity, suppressing the fluctuation beyond the classical limit.

Moreover, our formulation offers a highly transparent physical interpretation compared to most existing TURs and KURs in the quantum regime. While previous bounds often rely on the full density matrix $\rho$, require current information, or lack clear physical meaning, our precision limit is determined solely by the magnetizations, the system parameters $\omega$ and $\kappa$, and the particle number $N$ through the mean-field approximation. Finally, the global factor $N$ in both $B_{\mathrm{mb}}(\tau)$ and $B_{\mathrm{mb}}^{\mathrm{up}}(\tau)$ dictates a $1/N$ scaling of the fluctuation. This explicitly demonstrates that the enhanced precision is achieved through the cooperative nature of the many-body system.

 \paragraph{Numerical simulations.---}

\begin{figure*}[t]
\includegraphics[width=180mm]{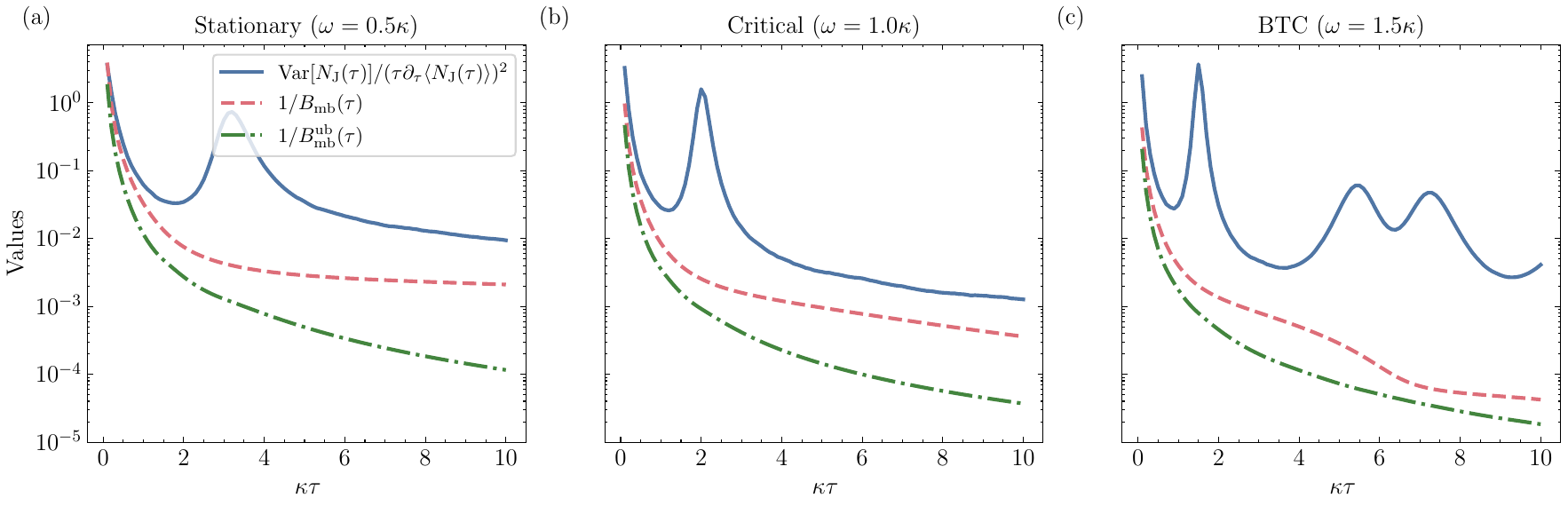}
\caption{\label{fig:tur_tau_dependence}Time dependence of the relative fluctuation and the theoretical lower bounds for a fixed particle number $N=100$ and $\kappa=1.0$. The relative fluctuation $\mathrm{Var}[N_\mathrm{J}(\tau)] / (\tau \partial_\tau \langle N_\mathrm{J}(\tau) \rangle)^2$ obtained from Monte Carlo simulations (blue solid lines) is plotted alongside the analytical bound $1/B_\mathrm{mb}(\tau)$ (red dashed lines) and the looser bound $1/B_\text{mb}^\text{ub}(\tau)$ (green dash-dotted lines) as a function of the measurement time $\kappa\tau$. The three panels correspond to the distinct dynamical phases: (a) the stationary phase with $\omega=0.5\kappa$, (b) the critical regime with $\omega=1.0\kappa$, and (c) the BTC phase with $\omega=1.5\kappa$. The system is initialized in a spin coherent state $|\theta, \phi\rangle=\exp[\theta(e^{i\phi}S_--e^{-i\phi}S_+)/2]|N/2, N/2\rangle$ with $\theta=0$ and $\phi=0$.
} 
\end{figure*}

To illustrate the validity and physical implications of the KUR derived in this study [\Eref{our_kur}], we perform numerical simulations of the continuously monitored collective spin model described by \Eref{lindblad}. We compare the relative fluctuation $\mathrm{Var}[N_\mathrm{J}(\tau)] / (\tau \partial_\tau \langle N_\mathrm{J}(\tau) \rangle)^2$ of the observable obtained from the quantum jump unraveling with our analytical lower bounds, $1/B_\mathrm{mb}(\tau)$ and $1/B_\text{mb}^\text{ub}(\tau)$, focusing on both their temporal evolution and system-size scaling across different dynamical phases. Here, the analytical bounds $1/B_\mathrm{mb}(\tau)$ and $1/B_\text{mb}^\text{ub}(\tau)$ are evaluated using Eqs. (\ref{B_mb}) and (\ref{B_mb_up}), respectively. The time-dependent magnetizations $m_\alpha(t)$ required for these expressions are obtained from the mean-field equations in \Eref{mean_field_equation}. We choose the spin coherent state \cite{maQuantumSpinSqueezing2011} as the initial state. We consider three representative parameter regimes of the driving strength $\omega$, namely stationary phase ($\omega=0.5\kappa$), the critical point ($\omega=1.0\kappa$), and the BTC phase ($\omega=1.5\kappa$). The dynamics of the continuously monitored open quantum system are simulated using the Monte Carlo method with 1000 trajectories to evaluate the relative fluctuation $\mathrm{Var}[N_\mathrm{J}(\tau)] / (\tau \partial_\tau \langle N_\mathrm{J}(\tau) \rangle)^2$ of the jump records.

We first examine the time dependence of the KUR. \FFref{fig:tur_tau_dependence} shows the relative fluctuation $\mathrm{Var}[N_\mathrm{J}(\tau)] / (\tau \partial_\tau \langle N_\mathrm{J}(\tau) \rangle)^2$ (blue solid line) alongside the analytical bounds $1/B_\mathrm{mb}(\tau)$ (red dashed line) and $1/B_\text{mb}^\text{ub}(\tau)$ (green dash dotted line) as a function of the measurement time $\tau$ for a fixed particle number $N=100$. Across all three dynamical phases including the stationary [\Fref{fig:tur_tau_dependence}(a)], critical [\Fref{fig:tur_tau_dependence}(b)], and BTC [\Fref{fig:tur_tau_dependence}(c)] regimes, the relative fluctuation from the Monte Carlo simulation is strictly bounded from below by the analytical expression $1/B_\mathrm{mb}(\tau)$. This result confirms the validity of our formulation. Furthermore, the mathematically simpler expression $1/B_\text{mb}^\text{ub}(\tau)$ correctly serves as a valid lower bound for $1/B_\mathrm{mb}(\tau)$. Comparing the different dynamical phases reveals an interesting property regarding the achievable precision. The actual precision obtained from the Monte Carlo simulation reaches its best value in the critical regime, followed by the BTC and stationary phases. On the other hand, the theoretical lower bound indicates that the ultimate precision limit has the potential to be most favorable in the BTC phase, followed by the critical and stationary regimes. It is worth noting that the oscillatory features characteristic of the BTC phase decay with increasing $\tau$ as a consequence of the finite system size $N$.

\begin{figure*}[t]
\includegraphics[width=180mm]{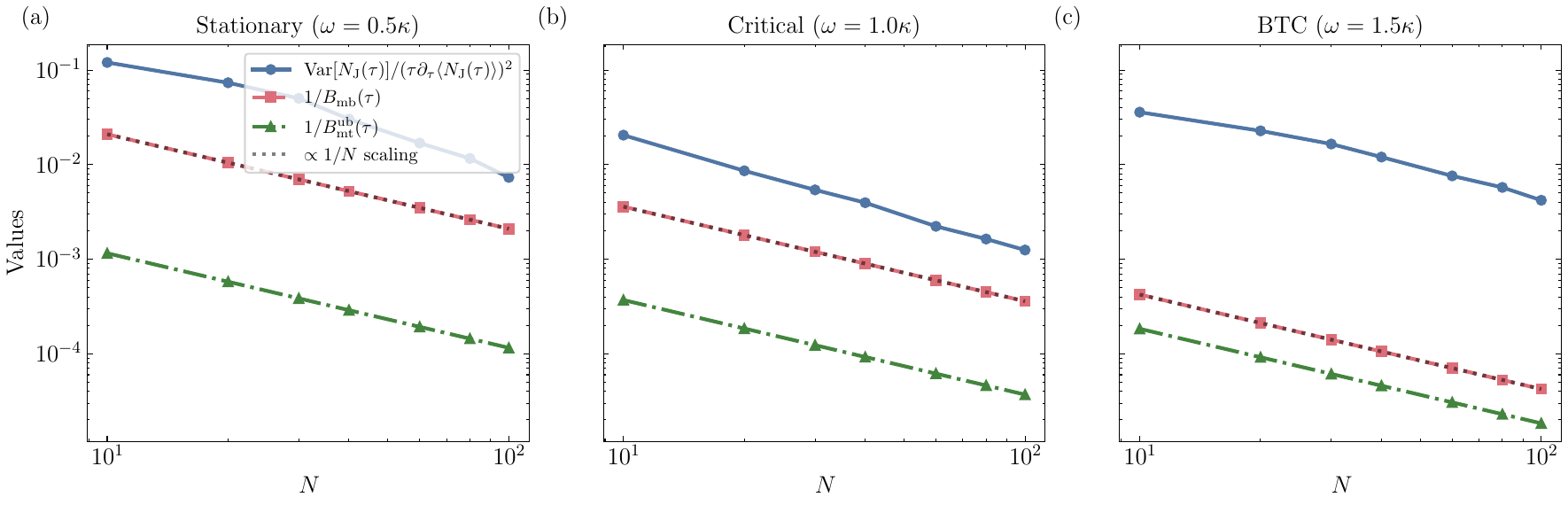}
\caption{\label{fig:tur_N_dependence}System size scaling of the relative fluctuation and the theoretical lower bounds at a fixed measurement time $\tau = 10.0$. The relative fluctuation $\mathrm{Var}[N_\mathrm{J}(\tau)]/(\tau \partial_\tau \langle N_\mathrm{J}(\tau) \rangle)^2$ obtained from Monte Carlo simulations (blue solid line with circles) is plotted alongside the analytical bound $1/B_\mathrm{mb}(\tau)$ (red dashed line with squares) and the looser bound $1/B_\mathrm{mb}^\mathrm{ub}(\tau)$ (green dash-dotted line with triangles) as a function of the particle number $N \in \{10, 20, 30, 40, 60, 80, 100\}$. The gray dotted lines indicate the $\propto 1/N$ scaling behavior. The three panels show the results for the distinct dynamical phases: (a) the stationary phase ($\omega = 0.5\kappa$), (b) the critical regime ($\omega = 1.0\kappa$), and (c) the BTC phase ($\omega = 1.5\kappa$). The initial state is the spin coherent state, with parameters identical to those used in Fig.~\ref{fig:tur_tau_dependence}.
} 
\end{figure*}

To explore the potential advantage of collective systems for precision, we investigate whether our setup exhibits a cooperative enhancement of precision. Specifically, to verify the scaling behavior predicted by our analytical formulation of $1/B_\mathrm{mb}(\tau)$ and $1/B_\text{mb}^\text{ub}(\tau)$, we examine the dependence of the KUR on the number of particles $N$.  \FFref{fig:tur_N_dependence} displays the $N$ dependence of the relative fluctuation and the theoretical lower bounds at a fixed measurement time $\tau=10.0$. In all parameter regimes, the Monte Carlo results and the analytical bounds follow the $1/N$ scaling indicated by the black dotted line. This scaling manifests the collective nature of the system. As the system size increases, the macroscopic classical activity and the quantum fluctuation term both grow, thereby suppressing the relative fluctuation by a factor of $1/N$. This agreement between the simulations and the theoretical lines underscores that our analytical bounds correctly capture the macroscopic features of the many-body system without requiring the full exponential computational cost of calculating the density matrix.

Finally, we discuss the tightness of the derived bounds. While our analytical lower bounds successfully capture the temporal dynamics and the $1/N$ collective scaling of the relative fluctuation, there exists a visible quantitative gap between the actual fluctuation of the jump observable $\mathrm{Var}[N_J(\tau)] / (\tau \partial_\tau \langle N_J(\tau) \rangle)^2$ and the theoretical limit $1/B_{\mathrm{mb}}(\tau)$ across all phases. This gap fundamentally originates from the nature of the quantum Cram\'{e}r-Rao bound. The analytical expression $B_{\mathrm{mb}}(\tau)$, which equals the exact quantum Fisher information $J(0)$, dictates the ultimate precision limit achievable by theoretically optimizing over all possible positive operator-valued measures (POVMs) on the system and its environment. Conversely, the observable $N_J(\tau)$ evaluated in our Monte Carlo simulations corresponds to a specific, fixed unraveling strategy, namely jump measurements. Extracting information solely from this jump record inherently results in a loss of information compared to the total information accessible via the optimal POVM.

Furthermore, this gap becomes most pronounced in the BTC phase [\Fref{fig:tur_tau_dependence}(c) and \Fref{fig:tur_N_dependence}(c)]. The BTC phase is dominated by macroscopic coherent oscillations driven by the Hamiltonian. While the quantum Fisher information incorporates the full precision potential of these coherent dynamics, the jump record is phase-insensitive. Consequently, jump measurements may become sub-optimal compared to phase-sensitive unravelings, resulting in a larger deviation from the theoretical bound.
We also note that the tightness of the bound is sensitive to the choice of the initial state. \paragraph{Conclusion.---}
In this study, we investigated the precision limits of many-body quantum systems with collective dissipation, which serves as a paradigmatic model for BTC, within the framework of the KUR. By employing a mean-field approximation that exploits the many-body nature of the system, we analytically derived rigorous lower bounds for the relative fluctuation of the time-integrated observable. These bounds are expressed solely through physically transparent quantities, such as magnetization and system parameters, providing a clear interpretation of the fundamental precision limits. Furthermore, we theoretically demonstrated a cooperative enhancement of precision, proving that collective effects suppress the uncertainty in proportion to the system size $N$.

Numerical simulations across the stationary, critical, and BTC phases validated these analytical findings. The results show that the actual precision reaches its maximum in the critical regime. Although the theoretical bounds indicate that the BTC phase has the potential for the highest precision. This agreement between theory and simulation underscores the ability of our analytical approach to capture many-body dynamics.

This work provides the first theoretical characterization of the precision limits in collective quantum many-body systems for an arbitrary particle number $N$. As the development of advanced quantum technologies scales up to macroscopic regimes, interacting many-body systems inherently serve as the fundamental physical basis for these architectures. Consequently, uncovering how genuine many-body effects redefine fundamental precision boundaries is a crucial theoretical objective. Our formulation clarifies this relationship, demonstrating how collective dynamics govern the fundamental limits of precision. This provides clear and interpretable guidelines for designing future quantum technologies that harness many-body phenomena. 

\paragraph{Acknowledgments.---}
This work was supported by JSPS KAKENHI Grant No. JP23K24915.

\end{document}


\setcounter{equation}{0}
\setcounter{figure}{0}
\setcounter{table}{0}
\setcounter{page}{1}
\makeatletter
\renewcommand{\theequation}{S\arabic{equation}}
\renewcommand{\thefigure}{S\arabic{figure}}
\renewcommand{\thetable}{S\arabic{table}}
\renewcommand{\thesection}{S\arabic{section}}
\makeatother

\preprint{APS/123-QED}

\title{Supplementary Material for \\``Kinetic Uncertainty Relation in Collective Dissipative Quantum Many-Body Systems''}

\author{Hayato Yunoki}
 \email{yunoki@biom.t.u-tokyo.ac.jp}

\affiliation{
 Department of Information and Communication Engineering,\\
 Graduate School of Information Science and Technology,\\
 The University of Tokyo, Tokyo 113-8656, Japan
}

\author{Yoshihiko Hasegawa}%
 \email{hasegawa@biom.t.u-tokyo.ac.jp}

\affiliation{Department of Electrical Engineering and Information Systems, Graduate School of Engineering, The University of Tokyo,
Tokyo 113-8656, Japan}

\newcommand{\refMainModel}{(2)}
\newcommand{\refMainBub}{(18)}
\newcommand{\refMainFs}{(19)}
\newcommand{\refMainFeff}{(20)}
\newcommand{\refMainBmb}{(15)}

\maketitle


This supplementary material describes the calculations introduced in the main text. The numbers of the equations and the figures are prefixed with S (e.g., Eq. (S1) or Fig. S1). Numbers without this prefix (e.g., Eq. (1) or Fig. 1) refer to items in the main text.

\section{Exact Formalism of the Quantum Dynamical Activity}
To analytically derive the many-body precision bounds $B_{\mathrm{mb}}(\tau)$ and $B_{\mathrm{mb}}^{\mathrm{ub}}(\tau)$ presented in the main text, our analysis begins with the exact formal expression of the quantum Fisher information at the unperturbed limit, $J(\theta=0)$. 

Let us consider a Lindblad master equation described by a system Hamiltonian $H$ and a set of jump operators $\{L_{m}\}$,
\begin{equation} \label{eq:S_generic_lindblad}
    \dot{\rho} = -i[H, \rho] + \sum_{m} \left( L_{m} \rho L_{m}^{\dagger} - \frac{1}{2} \{L_{m}^{\dagger} L_{m}, \rho\} \right) \equiv \mathbb{L}\rho,
\end{equation}
where we introduce the superoperator $\mathbb{L}$.

The quantity $J(0)$, recognized as the quantum dynamical activity, serves as the fundamental quantity in kinetic uncertainty relations and speed limit for quantum systems \cite{hasegawaUnifyingSpeedLimit2023}. For the dynamics governed by Eq.~\eqref{eq:S_generic_lindblad}, the exact analytical formulation of $J(0)$ has been established in recent studies \cite{nakajimaSymmetriclogarithmicderivativeFisherInformation2023, nishiyamaExactSolutionQuantum2024}. Here, we present the result derived in Ref.~\cite{nishiyamaExactSolutionQuantum2024}, which expresses $J(0)$ as
\begin{equation} \label{eq:S_exact_J0}
\begin{split}
    J(0) = \mathcal{A}(\tau) + 8 \int_{0}^{\tau} ds_{1} \int_{0}^{s_{1}} ds_{2} \mathrm{Re} \left( \mathrm{Tr} \left[ H_{\mathrm{eff}}^{\dagger} \tilde{H}(s_{1}-s_{2}) \rho(s_{2}) \right] \right) - 4 \left( \int_{0}^{\tau} ds \mathrm{Tr}[H \rho(s)] \right)^{2}.
\end{split}
\end{equation}
In this expression, $\mathcal{A}(\tau) = \int_{0}^{\tau} ds \sum_{m} \mathrm{Tr}[L_{m} \rho(s) L_{m}^{\dagger}]$ represents the classical dynamical activity, which corresponds to the average number of jumps. The operator $H_{\mathrm{eff}} = H - \frac{i}{2} \sum_{m} L_{m}^{\dagger} L_{m}$ denotes the non-Hermitian effective Hamiltonian, and $\tilde{H}(t) = e^{\mathbb{L}^{\dagger}t}H$ defines the time-evolved Hamiltonian in the Heisenberg picture with respect to the adjoint superoperator $\mathbb{L}^{\dagger}$, which acts on an arbitrary operator $\mathcal{O}$ as
\begin{equation} \label{eq:S_adjoint_superoperator}
    \mathbb{L}^{\dagger}\mathcal{O} = i[H, \mathcal{O}] + \sum_{m} \left( L_{m}^{\dagger} \mathcal{O} L_{m} - \frac{1}{2} \{L_{m}^{\dagger} L_{m}, \mathcal{O}\} \right).
\end{equation} 

Ref.~\cite{nishiyamaExactSolutionQuantum2024} also provides a theoretical upper bound for $J(0)$:
\begin{equation} \label{eq:S_upper_J0}
    J(0) \leq J^{\mathrm{ub}}(0) = \mathcal{A}(\tau) + 8 \int_{0}^{\tau} ds_{1} \sigma_{H}(s_{1}) \int_{0}^{s_{1}} ds_{2} \sigma_{H_{\mathrm{eff}}}(s_{2}).
\end{equation}
Here, the standard deviation for an arbitrary operator $\mathcal{O}$ at time $s$ is defined as $\sigma_{\mathcal{O}}(s) \equiv \sqrt{\langle \mathcal{O}^{\dagger} \mathcal{O} \rangle_s - |\langle \mathcal{O} \rangle_s|^{2}}$.

These exact expressions serve as the foundation of our analysis. In the following sections, we use Eqs.~\eqref{eq:S_exact_J0} and \eqref{eq:S_upper_J0} to derive the expression of $B_{\mathrm{mb}}(\tau)$ and $B_{\mathrm{mb}}^{\mathrm{ub}}(\tau)$.

\section{Derivation of the Upper Bound $B_{\mathrm{mb}}^{\mathrm{ub}}(\tau)$}

In this section, we derive the expression for $B_{\mathrm{mb}}^{\mathrm{ub}}(\tau)$ for the many-body system with collective dissipation described by Eq.~\refMainModel. Specifically, we evaluate $J^{\mathrm{ub}}(0)$ [Eq.~\eqref{eq:S_upper_J0}] for this model.

For our collective dissipative many-body system, the system Hamiltonian and the jump operator are given by
\begin{equation} \label{eq:S_model_operators}
    H = \omega S_x, \quad L = \sqrt{\frac{2\kappa}{N}} S_-,
\end{equation}
where $S^{\alpha} = \frac{1}{2}\sum_{i=1}^{N}\sigma^{(i)}_{\alpha}$ are the collective spin operators. The non-Hermitian effective Hamiltonian is constructed as
\begin{equation} \label{eq:S_Heff_model}
    H_{\mathrm{eff}} = H - \frac{i}{2} L^{\dagger}L = \omega S_x - i\frac{\kappa}{N} S_+S_-.
\end{equation}
We assume that the particle number $N$ is sufficiently large ($N \gg 1$) and apply the many-body mean-field approximation. Under the clustering assumption, the macroscopic state is characterized by the magnetization components $m_\alpha=\langle S_\alpha\rangle/(N/2)$ with a fixed total length $m_x^2+m_y^2+m_z^2=1$.

\subsection{Evaluation of the Classical Dynamical Activity $\mathcal{A}(\tau)$}
We first evaluate the classical part of the activity $\mathcal{A}(\tau) = \int_{0}^{\tau} ds \langle L^{\dagger}L \rangle_s = \frac{2\kappa}{N}\int_{0}^{\tau} ds \langle S_+ S_- \rangle_s$. 
Using the definition of the spin ladder operators $S_{\pm} = S_x \pm i S_y$, we can expand the product $S_+ S_-$ as
\begin{align} \label{eq:S_SplusSminus_expansion}
    S_+ S_- &= (S_x + i S_y)(S_x - i S_y) \nonumber \\
    &= S_x^2 + S_y^2 - i [S_x, S_y] \nonumber \\
    &= S_x^2 + S_y^2 + S_z.
\end{align}
where we used the commutation relation $[S_x, S_y] = i S_z$.
From the definition $S_\alpha = \sum_{i=1}^N s^i_\alpha$ with $s^i_\alpha = \sigma^i_\alpha / 2$, we decompose $S_x^2$ as
\begin{equation} \label{eq:S_Sx2_decomposition}
    \langle S_x^2 \rangle = \sum_{i=1}^N \langle (s^i_x)^2 \rangle + \sum_{i \neq j} \langle s^i_x s^j_x \rangle.
\end{equation}
The first term yields $\sum_{i=1}^N 1/4 = N/4$, as $(s^i_x)^2 = 1/4$. Under the mean-field approximation, we neglect correlations between different sites and factorize the second term as $\langle s_i^x s_j^x \rangle \approx \langle s_i^x \rangle \langle s_j^x \rangle = (m_x/2)^2$. Summing these contributions, we obtain
\begin{equation} \label{eq:S_Sx2_final}
    \langle S_x^2 \rangle \approx \frac{N}{4} + (N^2 - N) \frac{m_x^2}{4}.
\end{equation}
Following the same procedure for $\langle S_y^2 \rangle$ and substituting these into Eq.~\eqref{eq:S_SplusSminus_expansion}, we find
\begin{equation} \label{eq:S_transverse_squared_new}
    \langle S_x^2 + S_y^2 \rangle \approx \frac{N^2}{4}(m_x^2 + m_y^2) + \frac{N}{4}(2 - m_x^2 - m_y^2).
\end{equation}
The remaining term in Eq.~\eqref{eq:S_SplusSminus_expansion} is $\langle S_z \rangle = \frac{N}{2}m_z$, which is of order $O(N)$. In the thermodynamic limit ($N \gg 1$), the $O(N^2)$ terms in Eq.~\eqref{eq:S_transverse_squared_new} dominate, allowing us to neglect these $O(N)$ contributions. Consequently, using $m_x^2 + m_y^2 + m_z^2 = 1$, we arrive at
\begin{equation} \label{eq:S_mean_SplusSminus_final}
    \langle S_+ S_- \rangle \approx \frac{N^2}{4}(1 - m_z(s)^2).
\end{equation}
Substituting this into $\mathcal{A}(\tau)$ yields
\begin{equation} \label{eq:S_A_tau_evaluated}
    \mathcal{A}(\tau) \approx \frac{2\kappa}{N} \int_{0}^{\tau} ds \frac{N^2}{4}(1-m_z(s)^2) = \frac{\kappa N}{2} \int_{0}^{\tau} ds (1-m_z(s)^2).
\end{equation}

\subsection{Evaluation of the Standard Deviations}
Next, we evaluate the standard deviations $\sigma_{H}$ and $\sigma_{H_{\mathrm{eff}}}$. For the collective spin components $S_\alpha$, the variance $\mathrm{Var}(S_\alpha)$ can be calculated as done in Eq.~\eqref{eq:S_Sx2_decomposition}:
\begin{equation} \label{eq:S_Var_decomposition}
    \mathrm{Var}(S_\alpha) = \langle S_\alpha^2 \rangle - \langle S_\alpha \rangle^2 = \sum_{i=1}^N \langle (s^i_\alpha)^2 \rangle + \sum_{i \neq j} \langle s^i_\alpha s^j_\alpha \rangle - \left( \frac{N}{2}m_\alpha \right)^2.
\end{equation}
As shown in the previous section, the first term yields $N/4$ and the second term is factorized as $(N^2 - N)(m_\alpha/2)^2$ under the mean-field approximation. This leads to
\begin{equation} \label{eq:S_Var_final}
    \mathrm{Var}(S_\alpha) \approx \frac{N}{4} + \frac{N^2 - N}{4}m_\alpha^2 - \frac{N^2}{4}m_\alpha^2 = \frac{N}{4}(1 - m_\alpha^2).
\end{equation}

For the system Hamiltonian $H = \omega S_x$, the standard deviation is directly obtained from Eq.~\eqref{eq:S_Var_final} as
\begin{equation} \label{eq:S_sigma_H}
    \sigma_{H}(s) = \omega \sqrt{\mathrm{Var}(S_x)} = \omega \frac{\sqrt{N}}{2} \sqrt{1-m_x(s)^2} \equiv \frac{\sqrt{N}}{2} \mathcal{F}_S(s),
\end{equation}
which reproduces the definition of $\mathcal{F}_S(s)$ in Eq.~\refMainFs.

Applying the inequality for the standard deviation $\sigma_{A+B} \leq \sigma_{A} + \sigma_{B}$ to $H_{\mathrm{eff}} = \omega S_x - i\frac{\kappa}{N} S_+ S_-$, we can bound its fluctuation as
\begin{equation} \label{eq:S_sigma_Heff_inequality}
    \sigma_{H_{\mathrm{eff}}}(s) \leq \omega \sigma_{S_x}(s) + \frac{\kappa}{N} \sigma_{S_+ S_-}(s).
\end{equation}
To calculate $\sigma_{S_+ S_-}$, we first rewrite the product using the total spin operator $\mathbf{S}^2 = S_x^2 + S_y^2 + S_z^2$, which gives $S_+ S_- = \mathbf{S}^2 - S_z^2 + S_z$. 
Because the Hamiltonian and the collective jump operators preserve the permutation symmetry of the system, the dynamics are confined to the subspace of maximum total spin. In this subspace, $\mathbf{S}^2 = \frac{N}{2}\left(\frac{N}{2}+1\right)$ acts as a constant scalar. Since additive constants do not alter the variance, we obtain the relation:
\begin{equation} \label{eq:S_Var_exact_reduction}
    \mathrm{Var}(S_+ S_-) = \mathrm{Var}(-S_z^2 + S_z).
\end{equation}
To evaluate this variance, we employ the linearized mean-field approximation, decomposing the operator into its mean and the fluctuation: $S_z = \langle S_z \rangle + \delta S_z$. Substituting this into the operator yields
\begin{equation}
    -S_z^2 + S_z = -\langle S_z \rangle^2 - 2\langle S_z \rangle \delta S_z - (\delta S_z)^2 + \langle S_z \rangle + \delta S_z.
\end{equation}
Within the mean-field framework, we neglect the higher-order fluctuation term $(\delta S_z)^2$. Ignoring constant terms that do not contribute to the variance, we obtain
\begin{equation} \label{eq:S_Var_Sz_linearized}
    \mathrm{Var}(-S_z^2 + S_z) \approx (1 - 2\langle S_z \rangle)^2 \mathrm{Var}(S_z).
\end{equation}
Retaining only this leading-order term inside the squared parenthesis, the variance yields an $O(N^3)$ scaling:
\begin{align} \label{eq:S_Var_Sz2}
    \mathrm{Var}(S_+ S_-) &\approx 4\langle S_z \rangle^2 \mathrm{Var}(S_z) \nonumber \\
    &= 4 \left(\frac{N}{2}m_z\right)^2 \frac{N}{4}(1-m_z^2) = \frac{N^3}{4}m_z^2(1-m_z^2).
\end{align}
Taking the square root, we obtain $\sigma_{S_+ S_-} \approx \frac{N^{3/2}}{2}|m_z|\sqrt{1-m_z^2}$. Multiplying by the prefactor $\kappa/N$ yields
\begin{equation} \label{eq:S_sigma_dissipator}
    \frac{\kappa}{N} \sigma_{S_+ S_-}(s) \approx \kappa \frac{\sqrt{N}}{2} |m_z(s)|\sqrt{1-m_z(s)^2}.
\end{equation}
Substituting Eqs.~\eqref{eq:S_sigma_H} and \eqref{eq:S_sigma_dissipator} into Eq.~\eqref{eq:S_sigma_Heff_inequality}, we find the upper bound for the effective Hamiltonian's fluctuation:
\begin{equation} \label{eq:S_sigma_Heff_final}
    \sigma_{H_{\mathrm{eff}}}(s) \leq \frac{\sqrt{N}}{2} \left[ \omega \sqrt{1-m_x^2(s)} + \kappa |m_z|\sqrt{1-m_z^2(s)} \right] \equiv \frac{\sqrt{N}}{2} \mathcal{F}_{\mathrm{eff}}(s),
\end{equation}
which precisely corresponds to the function $\mathcal{F}_{\mathrm{eff}}(s)$ defined in Eq.~\refMainFeff. Crucially, both $\sigma_{H}$ and $\sigma_{H_{\mathrm{eff}}}$ scale as $O(\sqrt{N})$.

\subsection{Construction of $B_{\mathrm{mb}}^{\mathrm{ub}}(\tau)$}
Finally, we substitute the evaluated components into the formal upper bound Eq.~\eqref{eq:S_upper_J0}. 
\begin{align} \label{eq:S_J_ub_intermediate}
    J^{\mathrm{ub}}(0) &\leq \mathcal{A}(\tau) + 8 \int_{0}^{\tau} ds_1 \left( \frac{\sqrt{N}}{2}\mathcal{F}_S(s_1) \right) \int_{0}^{s_1} ds_2 \left( \frac{\sqrt{N}}{2}\mathcal{F}_{\mathrm{eff}}(s_2) \right) \nonumber \\
    &= \frac{\kappa N}{2} \int_{0}^{\tau} ds (1-m_z^2) + 2N \int_{0}^{\tau} ds_1 \mathcal{F}_S(s_1) \int_{0}^{s_1} ds_2 \mathcal{F}_{\mathrm{eff}}(s_2)\equiv B_{\mathrm{mb}}^{\mathrm{ub}}(\tau).
\end{align}

This rigorous inequality completes the derivation of the analytical upper bound $B_{\mathrm{mb}}^{\mathrm{ub}}(\tau)$ presented in Eq.~\refMainBub~of the main text. Notably, the classical dynamical activity and the quantum contribution term both grow linearly with $N$, demonstrating the cooperative enhancement of precision.

\section{Derivation of the Many-Body Quantum Dynamical Activity $B_{\mathrm{mb}}(\tau)$}

In this section, we derive the explicit expression for the quantum dynamical activity $B_{\mathrm{mb}}(\tau)$ [Eq.~\refMainBmb] for the many-body system with collective dissipation described by Eq.~\refMainModel. Specifically, we evaluate $J(0)$ [Eq.~\eqref{eq:S_exact_J0}] for this model under the mean-field approximation.

\subsection{Exact Expression and Covariance Formulation}
The exact expression for the dynamical activity $J(0)$ is given by
\begin{align} \label{eq:S_exact_J0_decomposition}
    J(0) &= \mathcal{A}(\tau) + 8 \int_{0}^{\tau} ds_1 \int_{0}^{s_1} ds_2 \mathrm{Re} \left[ \langle H_{\mathrm{eff}}^\dagger \tilde{H}(u) \rangle_{s_2} \right] - 4 \left( \int_{0}^{\tau} ds \langle H \rangle_s \right)^2,
\end{align}
where $u = s_1 - s_2$ and the expectation values are evaluated at the state $\rho(s_2)$. Using the symmetry of the integration domain, the last term can be rewritten as
\begin{equation} \label{eq:S_double_integral_mean}
    4 \left( \int_{0}^{\tau} ds \langle H \rangle_s \right)^2 = 8 \int_{0}^{\tau} ds_1 \int_{0}^{s_1} ds_2 \langle H \rangle_{s_1} \langle H \rangle_{s_2}.
\end{equation}

By combining this with the second term in Eq.~\eqref{eq:S_exact_J0_decomposition}, we can rigorously express the integrand in terms of a quantum covariance. To prove this, we first recall the definition of the covariance between the non-Hermitian effective Hamiltonian $H_{\mathrm{eff}}^\dagger$ and the time-evolved Hamiltonian $\tilde{H}(u)$ at the reference time $s_2$:
\begin{equation} \label{eq:S_covariance_definition}
    \mathrm{Cov}(H_{\mathrm{eff}}^\dagger, \tilde{H}(u)) = \langle H_{\mathrm{eff}}^\dagger \tilde{H}(u) \rangle_{s_2} - \langle H_{\mathrm{eff}}^\dagger \rangle_{s_2} \langle \tilde{H}(u) \rangle_{s_2}.
\end{equation}

Next, we evaluate the product of the expectation values in the second term. Since the effective Hamiltonian is given by $H_{\mathrm{eff}}^\dagger = H + \frac{i}{2} L^\dagger L$, its expectation value separates into a real and an imaginary part:
\begin{equation}
    \langle H_{\mathrm{eff}}^\dagger \rangle_{s_2} = \langle H \rangle_{s_2} + \frac{i}{2} \langle L^\dagger L \rangle_{s_2}.
\end{equation}
On the other hand, the expectation value of the evolved operator $\tilde{H}(u)$ at state $\rho(s_2)$ corresponds to the expectation value of $H$ at time $s_1$, i.e., $\langle \tilde{H}(u) \rangle_{s_2} = \mathrm{Tr}[\tilde{H}_S(u) \rho(s_2)] = \langle H \rangle_{s_1}$. Because $H$ is a Hermitian operator, its expectation value $\langle H \rangle_{s_1}$ is real.

Multiplying these two expectation values yields
\begin{equation}
    \langle H_{\mathrm{eff}}^\dagger \rangle_{s_2} \langle \tilde{H}(u) \rangle_{s_2} = \left( \langle H \rangle_{s_2} + \frac{i}{2} \langle L^\dagger L \rangle_{s_2} \right) \langle H \rangle_{s_1}.
\end{equation}
When taking the real part of this product, the term containing the imaginary coefficient $\frac{i}{2}$ multiplied by the real scalar $\langle H \rangle_{s_1}$ vanishes. Consequently, we obtain
\begin{equation} \label{eq:S_real_part_product}
    \mathrm{Re} \left[ \langle H_{\mathrm{eff}}^\dagger \rangle_{s_2} \langle \tilde{H}(u) \rangle_{s_2} \right] = \langle H \rangle_{s_2} \langle H \rangle_{s_1}.
\end{equation}

Therefore, by taking the real part of the covariance definition in Eq.~\eqref{eq:S_covariance_definition}, we arrive at the following:
\begin{align} \label{eq:S_exact_covariance_reduction}
    \mathrm{Re} \left[ \mathrm{Cov}(H_{\mathrm{eff}}^\dagger, \tilde{H}_S(u)) \right] &= \mathrm{Re} \left[ \langle H_{\mathrm{eff}}^\dagger \tilde{H}(u) \rangle_{s_2} \right] - \mathrm{Re} \left[ \langle H_{\mathrm{eff}}^\dagger \rangle_{s_2} \langle \tilde{H}(u) \rangle_{s_2} \right] \nonumber \\
    &= \mathrm{Re} \left[ \langle H_{\mathrm{eff}}^\dagger \tilde{H}(u) \rangle_{s_2} \right] - \langle H \rangle_{s_2} \langle H \rangle_{s_1}.
\end{align}
This allows us to elegantly reformulate the quantum dynamical activity as the integral of the real part of the covariance:
\begin{equation} \label{eq:S_J0_covariance_form}
    J(0) = \mathcal{A}(\tau) + 8 \int_{0}^{\tau} ds_1 \int_{0}^{s_1} ds_2 \mathrm{Re} \left[ \mathrm{Cov}(H_{\mathrm{eff}}^\dagger, \tilde{H}(u)) \right].
\end{equation}

Substituting $H = \omega S_x$ and $H_{\mathrm{eff}}^\dagger = \omega S_x + i\frac{\kappa}{N}S_+S_-$, the covariance term strictly separates into two components based on the property $\mathrm{Re}[iZ] = -\mathrm{Im}[Z]$:
\begin{equation} \label{eq:S_Cov_separated}
    \mathrm{Re} \left[ \mathrm{Cov}(H_{\mathrm{eff}}^\dagger, \tilde{H}(u)) \right] = \omega^2 \mathrm{Re} \left[ \mathrm{Cov}(S_x, \tilde{S}_x(u)) \right] - \frac{\kappa \omega}{N} \mathrm{Im} \left[ \mathrm{Cov}(S_+ S_-, \tilde{S}_x(u)) \right].
\end{equation}

\subsection{Linearized Dynamics of the Operators}
We determine the time evolution of the operator $\tilde{S}_\alpha(u)$ (where $u = s_1 - s_2$) in the Heisenberg picture. For our model, the evolution is governed by the adjoint superoperator $\mathcal{L}^\dagger$:
\begin{equation} \label{eq:S_adjoint_master}
    \frac{d}{du} \tilde{A}(u) = \mathcal{L}^\dagger[\tilde{A}(u)] = i[\omega S_x, \tilde{A}(u)] + \frac{2\kappa}{N} \left( S_+ \tilde{A}(u) S_- - \frac{1}{2}\{ S_+ S_-, \tilde{A}(u) \} \right).
\end{equation}

The formal solution for the evolved operator is $\tilde{A}(u) = \mathcal{V}^\dagger(u)[A]$, where $\mathcal{V}^\dagger(u) = \mathcal{T} \exp(\int_0^u \mathcal{L}^\dagger dt)$ is the adjoint dynamical superoperator. Crucially, the time evolution $\mathcal{V}^\dagger(u)$ and the generator $\mathcal{L}^\dagger$ are strictly commutative. This mathematical property allows us to rearrange the derivative as:
\begin{equation}
    \frac{d}{du} \tilde{S}_\alpha(u) = \frac{d}{du} \mathcal{V}^\dagger(u)[S_\alpha] = \mathcal{L}^\dagger \Bigl[ \mathcal{V}^\dagger(u)[S_\alpha] \Bigr] = \mathcal{V}^\dagger(u) \Bigl[ \mathcal{L}^\dagger[S_\alpha] \Bigr].
\end{equation}
This strictly justifies calculating the action of $\mathcal{L}^\dagger$ on the unevolved operators $S_\alpha$ without tildes, and applying the time evolution $\mathcal{V}^\dagger(u)$ at the end.

We now explicitly evaluate $\mathcal{L}^\dagger[S_\alpha]$ for $\alpha \in \{x, y, z\}$.
\begin{itemize}
    \item For $\alpha = x$: The unitary term vanishes ($i[\omega S_x, S_x] = 0$). For the dissipator $\mathcal{D}^\dagger[S_x]=\frac{2\kappa}{N}(S_+S_xS_--\frac{1}{2}\{S_+S_-, S_x\})$, using $[S_x, S_-] = S_z$, we rewrite the first term as $S_+ S_x S_- = S_+ S_- S_x + S_+ S_z$. Expanding the anticommutator yields $\{S_+ S_-, S_x\} = 2 S_+ S_- S_x + [S_x, S_+ S_-]$. With $[S_x, S_+ S_-] = -S_z S_- + S_+ S_z$, the dissipator simplifies to $\mathcal{D}^\dagger[S_x] = \frac{\kappa}{N} (S_+ S_z + S_z S_-)$.
    \item For $\alpha = y$: The unitary term is $i[\omega S_x, S_y] = -\omega S_z$. For the dissipator, using $[S_y, S_-] = -iS_z$ and $[S_y, S_+ S_-] = -iS_z S_- - iS_+ S_z$, we rigorously obtain $\mathcal{D}^\dagger[S_y] = -i\frac{\kappa}{N}(S_+ S_z - S_z S_-)$.
    \item For $\alpha = z$: The unitary term is $i[\omega S_x, S_z] = \omega S_y$. For the dissipator, using $[S_z, S_-] = -S_-$ and the fact that $[S_z, S_+ S_-] = 0$, we find $\mathcal{D}^\dagger[S_z] = -\frac{2\kappa}{N} S_+ S_-$.
\end{itemize}

We apply the mean-field approximation $AB \approx \langle A \rangle B + A \langle B \rangle - \langle A \rangle \langle B \rangle$. Substituting $\langle S_\pm \rangle = \frac{N}{2}(m_x \pm i m_y)$ and $\langle S_z \rangle = \frac{N}{2}m_z$, we linearize the results:
\begin{itemize}
    \item For $\alpha = x$: $S_+ S_z + S_z S_- \approx N(m_x S_z + m_z S_x) - \text{const.}$, yielding $\mathcal{L}^\dagger[S_x] \approx \kappa(m_z S_x + m_x S_z) + c_x$.
    \item For $\alpha = y$: $-i(S_+ S_z - S_z S_-) \approx N(m_y S_z + m_z S_y) - \text{const.}$, yielding $\mathcal{L}^\dagger[S_y] \approx \kappa(m_z S_y + m_y S_z) - \omega S_z + c_y$.
    \item For $\alpha = z$: $S_+ S_- \approx N(m_x S_x + m_y S_y) - \text{const.}$, yielding $\mathcal{L}^\dagger[S_z] \approx -2\kappa m_x S_x + (-2\kappa m_y + \omega) S_y + c_z$.
\end{itemize}

This yields a closed system of linear differential equations:
\begin{equation} \label{eq:S_linear_diff_eq}
    \frac{d}{du} \begin{pmatrix} \tilde{S}_x \\ \tilde{S}_y \\ \tilde{S}_z \end{pmatrix} = K(s_2+u) \begin{pmatrix} \tilde{S}_x \\ \tilde{S}_y \\ \tilde{S}_z \end{pmatrix} + \vec{c},
\end{equation}
where $\vec{c}$ is a vector of scalar constants and $K$ is defined as:
\begin{equation} \label{eq:S_Jacobian_matrix_correct}
K(t) = \begin{pmatrix}
\kappa m_z(t) & 0 & \kappa m_x(t) \\
0 & \kappa m_z(t) & \kappa m_y(t) - \omega \\
-2\kappa m_x(t) & -2\kappa m_y(t) + \omega & 0
\end{pmatrix}.
\end{equation}
By solving this differential equation, the time-evolved operator can be expressed as 
\begin{equation}
    \tilde{S}_x(u) = \sum_{\alpha \in \{x,y,z\}} U_{x\alpha}(s_1, s_2) S_\alpha + C_x(u),
\end{equation}
where $\tilde{S}_\alpha(u=0)=S_\alpha$, $U(s_1, s_2) = \mathcal{T} \exp \left( \int_{s_2}^{s_1} dt' K(t') \right)$ is the time-ordered propagator, and $C_x(u)$ is a scalar function arising from the integration of the constant vector $\vec{c}$. 

Because the covariance between any operator and a scalar constant is identically zero ($\mathrm{Cov}(A, C_x(u)) = 0$), the integrated constant term $C_x(u)$ completely drops out. Furthermore, since the matrix $K(t)$ consists solely of real parameters, the propagator elements $U_{x\alpha}(s_1, s_2)$ are strictly real numbers. Therefore, taking the real and imaginary parts of the covariance reduces to:
\begin{align}
    \mathrm{Re} \left[ \mathrm{Cov}(S_x, \tilde{S}_x(u)) \right] &= \sum_{\alpha} U_{x\alpha}(s_1, s_2) \mathrm{Re} \left[ \mathrm{Cov}(S_x, S_\alpha) \right], \label{eq:S_Cov_Re_expansion} \\
    \mathrm{Im} \left[ \mathrm{Cov}(S_+ S_-, \tilde{S}_x(u)) \right] &= \sum_{\alpha} U_{x\alpha}(s_1, s_2) \mathrm{Im} \left[ \mathrm{Cov}(S_+ S_-, S_\alpha) \right]. \label{eq:S_Cov_Im_expansion}
\end{align}

\subsection{Evaluation of the Covariances}
We analytically evaluate covariances at state $\rho(s_2)$ by rigorously separating the site summations into single-site ($i=j$) and two-site ($i \neq j$) contributions. Using $S_\mu = \sum_{i=1}^N s_\mu^i$ with $s_\mu^i = \sigma_\mu^i / 2$, the second moment expands as
\begin{equation}
    \langle S_x S_\alpha \rangle = \sum_{i \neq j} \langle s_x^i s_\alpha^j \rangle + \sum_{i=1}^N \langle s_x^i s_\alpha^i \rangle.
\end{equation}
Since the state is approximated by a factorized state in the mean-field limit, inter-site correlations vanish, yielding $\langle s_x^i s_\alpha^j \rangle = \langle s_x^i \rangle \langle s_\alpha^j \rangle = \frac{1}{4}m_x m_\alpha$.
For the single-site term, we use the exact Pauli algebra $s_x^i s_\alpha^i = \frac{1}{4} \delta_{x\alpha} I + \frac{i}{4} \sum_\lambda \epsilon_{x\alpha\lambda} \sigma_\lambda^i$. Summing over $N$ sites gives $\frac{N}{4} \delta_{x\alpha} + i \frac{N}{4} \sum_\lambda \epsilon_{x\alpha\lambda} m_\lambda$.
Using the product of the means $\langle S_x \rangle \langle S_\alpha \rangle = \frac{N^2}{4} m_x m_\alpha$, the covariance becomes
\begin{equation} \label{eq:S_Exact_Covariance}
\begin{split}
    \mathrm{Cov}(S_x, S_\alpha) =&  \left( \frac{N^2}{4} m_x m_\alpha - \frac{N}{4} m_x m_\alpha + \frac{N}{4} \delta_{x\alpha} + i\frac{N}{4} \sum_\lambda \epsilon_{x\alpha\lambda} m_\lambda \right) - \frac{N^2}{4} m_x m_\alpha \\
    =&\frac{N}{4} (\delta_{x\alpha} - m_x m_\alpha) + i \frac{N}{4} \sum_\lambda \epsilon_{x\alpha\lambda} m_\lambda.
\end{split}
\end{equation}
Taking the real part eliminates the imaginary commutator terms:
\begin{equation} \label{eq:S_Re_Cov_evaluated}
    \mathrm{Re} \left[ \mathrm{Cov}(S_x, S_\alpha) \right] = \frac{N}{4} (\delta_{x\alpha} - m_x m_\alpha).
\end{equation}

For the imaginary part of the covariance, we first derive its relation to the commutator. By definition, the covariance is $\mathrm{Cov}(A, B) = \langle AB \rangle - \langle A \rangle \langle B \rangle$. Here, both $A = S_+ S_-$ and $B = S_\alpha$ are Hermitian operators. Because the expectation value of any Hermitian operator is real, the product of their means $\langle A \rangle \langle B \rangle$ is a real number. Therefore, the imaginary part of the covariance is determined by $\mathrm{Im}[\mathrm{Cov}(A, B)] = \mathrm{Im}[\langle AB \rangle]$. 
For any complex number $Z$, its imaginary part is given by $(Z - Z^*)/(2i)$. Setting $Z = \langle AB \rangle$, its complex conjugate is $Z^* = \langle (AB)^\dagger \rangle = \langle B^\dagger A^\dagger \rangle = \langle BA \rangle$. This leads to
\begin{equation} \label{eq:S_Im_Cov_commutator_proof}
    \mathrm{Im}[\mathrm{Cov}(A, B)] = \frac{1}{2i} \left( \langle AB \rangle - \langle BA \rangle \right) = \frac{1}{2i} \langle [A, B] \rangle.
\end{equation}

Using $S_+ S_- = \mathbf{S}^2 - S_z^2 + S_z$, we calculate the commutators $[S_+S_-, S_\alpha]$ for each component $\alpha \in \{x, y, z\}$.

For $\alpha = x$, 
using $[\mathbf{S}^2, S_x]=0$, $[S_z, S_x]=iS_y$, and $[-S_z^2, S_x] = -i(S_y S_z + S_z S_y)$, we obtain
\begin{equation}
    [S_+ S_-, S_x] = [\mathbf{S}^2 - S_z^2 + S_z, S_x] = -i(S_y S_z + S_z S_y) + iS_y.
\end{equation}
To evaluate the expectation value $\langle S_y S_z + S_z S_y \rangle$, we perform the site separation. The single-site contribution vanishes due to the anti-commutation relation of Pauli matrices, $\{s_y^i, s_z^i\} = 0$. The two-site contribution factorizes as $\sum_{i \neq j} 2 \langle s_y^i \rangle \langle s_z^j \rangle = 2 N(N-1) \frac{m_y}{2} \frac{m_z}{2} = \frac{N^2-N}{2}m_y m_z$. Retaining the leading $O(N^2)$ term, we find
\begin{equation} \label{eq:S_Im_Cov_Sx}
    \mathrm{Im} \left[ \mathrm{Cov}(S_+ S_-, S_x) \right] \approx \frac{1}{2i} \left( -i \frac{N^2}{2} m_y m_z \right) = -\frac{N^2}{4} m_y m_z.
\end{equation}

For $\alpha = y$,
using $[S_z, S_y]=-iS_x$ and $[-S_z^2, S_y] = i(S_x S_z + S_z S_x)$, the commutator expands as
\begin{equation}
    [S_+ S_-, S_y] = [\mathbf{S}^2 - S_z^2 + S_z, S_y] = i(S_x S_z + S_z S_x) - iS_x.
\end{equation}
Applying the same site separation, the single-site terms for $S_x S_z + S_z S_x$ vanish because $\{s_x^i, s_z^i\} = 0$. The two-site terms factorize to yield $\langle S_x S_z + S_z S_x \rangle \approx \frac{N^2}{2} m_x m_z$ in the leading order. Thus, the imaginary part is evaluated as
\begin{equation} \label{eq:S_Im_Cov_Sy}
    \mathrm{Im} \left[ \mathrm{Cov}(S_+ S_-, S_y) \right] \approx \frac{1}{2i} \left( i \frac{N^2}{2} m_x m_z \right) = \frac{N^2}{4} m_x m_z.
\end{equation}

For $\alpha = z$,
since $\mathbf{S}^2$, $S_z^2$, and $S_z$ commute with $S_z$,
\begin{equation}
    [S_+ S_-, S_z] = [\mathbf{S}^2 - S_z^2 + S_z, S_z] = 0.
\end{equation}
Consequently, the covariance vanishes: $\mathrm{Im} \left[ \mathrm{Cov}(S_+ S_-, S_z) \right] = 0$.

\vspace{1em}
Substituting these explicitly evaluated values for $\alpha \in \{x, y, z\}$ into the imaginary covariance expansion Eq.~\eqref{eq:S_Cov_Im_expansion}, the term neatly simplifies to
\begin{equation} \label{eq:S_Im_Cov_evaluated}
    -\frac{\kappa \omega}{N} \mathrm{Im} \left[ \mathrm{Cov}(S_+ S_-, \tilde{S}_x(u)) \right] = \frac{\kappa \omega N}{4} m_z (m_y U_{xx} - m_x U_{xy}).
\end{equation}

\subsection{Final Analytical Expression for $B_{\mathrm{mb}}(\tau)$}
Finally, we substitute the analytically evaluated Eq.~\eqref{eq:S_Re_Cov_evaluated} and Eq.~\eqref{eq:S_Im_Cov_evaluated} into the exact formula Eq.~\eqref{eq:S_J0_covariance_form}. Together with the classical activity $\mathcal{A}(\tau)$ derived previously, we arrive at the final analytical expression for the many-body quantum dynamical activity:
\begin{align} \label{eq:S_B_mb_final_result}
    B_{\mathrm{mb}}(\tau) &= \frac{\kappa N}{2} \int_{0}^{\tau} ds (1 - m_z(s)^2) \nonumber \\
    &\quad + 2N \omega \int_{0}^{\tau} ds_1 \int_{0}^{s_1} ds_2 \Biggl[ \omega \sum_{\alpha \in \{x,y,z\}} U_{x\alpha}(s_1, s_2) \left( \delta_{x\alpha} - m_x(s_2)m_\alpha(s_2) \right) \nonumber \\
    &\quad + \kappa m_z(s_2) \left( m_y(s_2) U_{xx}(s_1, s_2) - m_x(s_2) U_{xy}(s_1, s_2) \right) \Biggr].
\end{align}

\section{Numerical Verification of the Analytical Expressions}
To confirm the validity of our analytical derivations, we numerically evaluate and compare the exact quantum Fisher information $J(0)$, the analytical many-body quantum dynamical activity $B_{\mathrm{mb}}(\tau)$, and its upper bound $B_{\mathrm{mb}}^{\mathrm{ub}}(\tau)$.

We consider a system size of $N = 40$ spins. The system is prepared in a spin coherent state pointing in the positive $y$-direction, denoted by $|\psi(0)\rangle = |\theta=\pi/2, \phi=\pi/2\rangle$ \cite{maQuantumSpinSqueezing2011}. This corresponds to the initial magnetization vector $m(0) = (0, 1, 0)$. 

The exact quantum dynamical activity (equivalent to the quantum Fisher information $J(0)$) is computed directly from its definition using the full quantum master equation, without relying on any approximations. Conversely, the analytical expression $B_{\mathrm{mb}}(\tau)$ and the upper bound $B_{\mathrm{mb}}^{\mathrm{ub}}(\tau)$ are evaluated by mean-field equations for the trajectory $\vec{m}(t)$.

\begin{figure}[t]
    \centering
    \includegraphics[width=\linewidth]{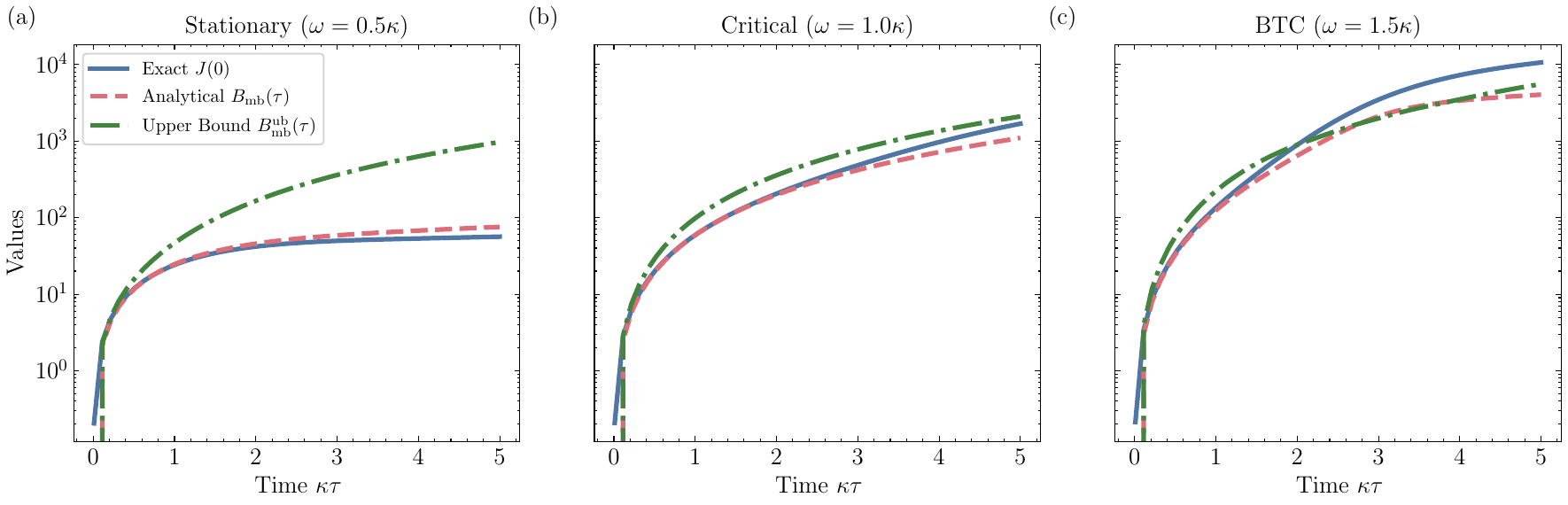}
    \caption{Numerical verification of the many-body quantum dynamical activity $B_{\mathrm{mb}}(\tau)$ and its upper bound $B_{\mathrm{mb}}^{\mathrm{ub}}(\tau)$. The exact quantum Fisher information $J(0)$ (blue solid line) is compared with the derived analytical expression $B_{\mathrm{mb}}(\tau)$ (red dashed line) and the upper bound $B_{\mathrm{mb}}^{\mathrm{ub}}(\tau)$ (green dash-dotted line) for a system of $N=40$ spins. The three panels show different dynamical phases: (a) the stationary phase ($\omega = 0.5\kappa$), (b) the critical point ($\omega = 1.0\kappa$), and (c) the boundary time crystal (BTC) phase ($\omega = 1.5\kappa$). The system is initially prepared in a spin coherent state pointing in the $y$-direction, $m(0) = (0, 1, 0)$.}
    \label{fig:S_time_evolution}
\end{figure}

We investigate three representative dynamical phases: the stationary phase ($\omega = 0.5\kappa$), the critical point ($\omega = 1.0\kappa$), and the boundary time crystal (BTC) phase ($\omega = 1.5\kappa$). As shown in Fig.~\ref{fig:S_time_evolution}, the analytical result $B_{\mathrm{mb}}(\tau)$ exhibits agreement with the exact $J(0)$ across all time scales and dynamical phases. This numerical coincidence strongly justifies the algebraic procedures detailed in the previous sections. Furthermore, the analytically derived upper bound $B_{\mathrm{mb}}^{\mathrm{ub}}(\tau)$ confirm its validity as a reliable bound.

%